# Change of pairing symmetry in the iron-based superconductor KFe$_2$As$_2$

F. F. Tafti[1],  A. Juneau-Fecteau[1], M.-È. Delage[1], S. René de Cotret[1], J.-Ph. Reid[1],
A. F. Wang[2], X.-G. Luo[2], X. H. Chen[2], N. Doiron-Leyraud[1] & Louis Taillefer[1,3]

*1 Département de physique & RQMP, Université de Sherbrooke, Sherbrooke, Québec J1K 2R1, Canada*

*2 Hefei National Laboratory for Physical Sciences at Microscale & Department of Physics, University of Science and Technology of China, Hefei, Anhui 230026, China*

*3 Canadian Institute for Advanced Research, Toronto, Ontario M5G 1Z8, Canada*

**The pairing mechanism in iron-based superconductors is the subject of ongoing debate[1,2,3,4]. Proximity to an antiferromagnetic phase suggests that pairing is mediated by spin fluctuations, but orbital fluctuations have also been invoked[5]. The former typically favour a pairing state of extended $s$-wave symmetry with a gap that changes sign between electron and hole Fermi surfaces[6,7,8,9] ($s_\pm$), while the latter yield a standard $s$-wave state without sign change[5] ($s_{++}$). Here we show that applying pressure to KFe$_2$As$_2$ induces a change of pairing state. The critical temperature $T_c$ decreases with pressure initially, and then suddenly increases, above a critical pressure $P_c$. The constancy of the Hall coefficient through $P_c$ rules out a change in the Fermi surface. There is compelling evidence that the pairing state below $P_c$ is $d$-wave, from bulk measurements at ambient pressure[10,11,12,13,14]. Above $P_c$, the high sensitivity to disorder argues for a particular kind of $s_\pm$ state[15]. The change from $d$-wave to $s$-wave is likely to proceed via an unusual $s + id$ state that breaks time-reversal symmetry[16,17,18]. The proximity of two distinct pairing states found here experimentally is natural given the near degeneracy of $d$-wave and $s_\pm$ states found**

**theoretically[19,20,21,22]. These findings make a compelling case for spin-fluctuation-mediated superconductivity in this key iron-arsenide material.**

$KFe_2As_2$ is a stoichiometric iron arsenide with a superconducting critical temperature $T_c = 4$ K. It is a member of the extensively studied "122" family of iron-based superconductors[23]. Single crystals can be grown with very high purity, making it by far the cleanest of the iron-based superconductors. Its high hole concentration is such that its Fermi surface does not contain the usual electron pocket at the $X$ point (of the unfolded Brillouin zone); it consists mainly of three hole-like cylinders: two located at the zone center ($\Gamma$) and one at the corner ($M$) (Fig. 1a). There is no antiferromagnetic order, but there are antiferromagnetic spin fluctuations, detected by inelastic neutron scattering[24].

In iron-based superconductors, spin fluctuations generally favour the $s_{\pm}$ pairing state in which the gap changes sign between hole and electron pockets[1,2,3,4] (Fig. 1b). In the absence of the electron pocket at $X$, this mechanism becomes much less effective, and functional-renormalization-group calculations show that a $d$-wave state (Fig. 1c) is the most stable state in $KFe_2As_2$ (ref. 19). Other theoretical methods find that $s_{\pm}$ and $d$-wave are very close in energy[20,21]. Experimentally, thermal conductivity studies in $KFe_2As_2$ make a compelling case for $d$-wave symmetry[10,11,12,13]: line nodes are found to be vertical and present on all Fermi surfaces, and the thermal conductivity is independent of impurity scattering, as expected of symmetry-imposed line nodes[25]. A $d$-wave state is also consistent with penetration depth data[14]. However, in a recent ARPES study of $KFe_2As_2$, vertical line nodes in the gap were observed on only one of the three Fermi surfaces[26]. To explain this, a particular kind of $s_{\pm}$ state was proposed[15] where the sign change is between the two $\Gamma$-centered hole pockets (Fig. 1d).

To help clarify the situation, we have studied the effect of pressure on $KFe_2As_2$, by measuring the resistivity and Hall effect in two single crystals, labeled sample A and sample B (Table S1). As seen in Fig. 2a, we find that $T_c$ decreases initially with pressure, but then, at a critical pressure $P_c = 17.5$ kbar, it suddenly starts to rise. $T_c$ varies linearly on either side of $P_c$, producing a V-shaped dependence of $T_c$ on $P$. This sharp inversion in the effect of pressure on $T_c$ is our central finding, reproduced in both samples (Fig. S1). There are two possible mechanisms: 1) a Lifshitz transition, whereby the Fermi surface undergoes a sudden change, or 2) a phase transition with broken symmetry. In Fig. 3a, we see that the Hall coefficient $R_H$ in the $T = 0$ limit remains completely unchanged by pressure, right through $P_c$ (Fig. 3b). A Lifshitz transition, such as the appearance of an electron pocket, would produce a sudden change in $R_H(0)$. It can therefore be excluded as a possible cause for the rise of $T_c$ beyond $P_c$.

We deduce that a phase transition occurs at $P_c$. Any density-wave or structural transition that breaks translational symmetry would reconstruct the Fermi surface, and cause associated anomalies in the transport properties. Such transitions are therefore ruled out by the absence of any anomaly in $R_H$ (Fig. 3b) and in the electrical resistivity $\rho$ (Fig. 3c) at $P_c$. Note that density-wave phases such as the antiferromagnetic phase in $Ba_{1-x}K_xFe_2As_2$ with $x < 0.4$ generally compete with superconductivity and so produce a dome-shaped curve of $T_c$ vs $P$ or $x$ (refs. 1, 2, 12), not a V-shaped curve as seen here (Fig. 2a). We conclude that what occurs at $P_c$ is not a transition in the normal-state electronic properties, but a transition to a superconducting phase of a different symmetry.

For the phase above $P_c$, the effect of impurity scattering on $T_c$ rules out the standard $s_{++}$ state. At ambient pressure, 4% Co impurities in $KFe_2As_2$ suppress $T_c$ to zero[13] – the critical value of the residual resistivity being $\rho_0{}^{crit} = 4.5$ μΩ cm (refs. 11, 13). This is consistent with a $d$-wave state, whose $T_c$ is expected to vanish when the scattering rate is of the order of $T_c$ (ref. 11). We measured the resistivity of a sample of $KFe_2As_2$ with 3.4% Co impurities, in which $\rho_0 = 3.8$ μΩ cm and $T_c = 1.7$ K at ambient pressure (Table S1 and Fig. S3). Under pressure, the $T_c$ of this Co-doped sample is suppressed to zero and does not re-emerge above $P_c$ (Fig. 4). This shows that the superconducting state above $P_c$ cannot be $s_{++}$, a state that is insensitive to non-magnetic impurities. Neither could this high-pressure state be the same $s_\pm$ state as in the usual $BaFe_2As_2$-based superconductors (with an electron pocket at $X$), since $T_c$ in these materials is very robust against impurity scattering[27].

A plausible candidate for the phase above $P_c$ is the $s_\pm$ state proposed in ref. 15 (for ambient pressure), with a sign change between the two hole pockets at $\Gamma$ (Fig. 1d). Given the similarity of these two pockets, inter-band scattering is likely to be significant and $T_c$ is therefore expected to be rather sensitive to disorder, as observed. The fact that, above $P_c$, $T_c$ rises even though $\rho$ continues to decrease (Fig. 3c) is consistent with calculations for this type of $s_\pm$ state, which require that the interaction between fermions be largest at small momentum transfer[15]. As mentioned above, this type of $s_\pm$ state is consistent with the ARPES study that finds a large, angle-dependent gap on the two $\Gamma$-centered hole pockets and a small nodeless gap on the $M$-centered hole pocket[26]. We suggest that the state measured by ARPES at ambient pressure is this $s_\pm$ state, stabilized at the cleaved (polar) surface of $KFe_2As_2$ even though the bulk is in a $d$-wave state.

Our pressure study of $KFe_2As_2$ reveals a close proximity between two different pairing states. To our knowledge, this is the first instance of a superconductor in which two distinct but intersecting domes of superconductivity with different symmetries are revealed by the application of pressure. When $d$-wave and $s$-wave phases come together in this way, an intermediate $s + id$ phase that breaks time-reversal symmetry is likely to intervene[16,17,18,22,28]. One signature of such a phase is a spontaneous internal magnetic field that appears below $T_c$, which could in principle be detected with muons, in clean samples of KFe$_2$As$_2$ at pressures near $P_c$.

In iron-based superconductors with a Fermi surface that contains hole and electron pockets (Fig. 1b), the natural proximity of $s_\pm$ and $d$-wave states was nicely revealed by calculations[22] where the strength of the $(\pi, \pi)$ spin fluctuations (connecting two nearby electron pockets) was gradually increased for a fixed strength of $(\pi, 0)$ spin fluctuations (connecting hole and electron pockets). A V-shaped variation of $T_c$ is obtained, as the superconducting phase goes from $s_\pm$ to $d$-wave. Impurity scattering suppresses $T_c$ on both sides of the transition and opens up an intermediate region without superconductivity[22]. This is the phenomenology we observe in KFe$_2$As$_2$, where an analogous mechanism of competing interactions may be at play even if the Fermi surface does not contain an electron pocket.

Chemical substitution is another tuning parameter, complementary to pressure. Substituting K for Ba in KFe$_2$As$_2$, for example, will also cause a change of pairing symmetry. Indeed, the superconducting gap in K$_x$Ba$_{1-x}$Fe$_2$As$_2$ at $x = 0.4$ does not contain any nodes, so it cannot be a $d$-wave gap[12]. Therefore, a transition from $d$-wave to some other pairing symmetry, most likely $s$-wave, will inevitably occur between $x = 1.0$ and $x = 0.4$, with an associated ten-fold enhancement of $T_c$ (ref. 12).

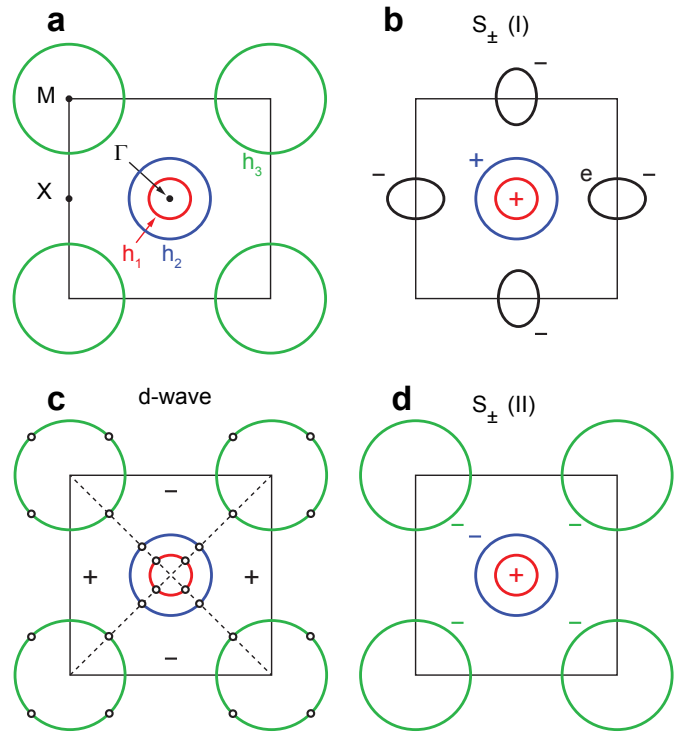

**Figure 1 | Fermi surface of KFe$_2$As$_2$ and possible superconducting states.**

**a)** Schematic sketch of the main Fermi surface sheets of KFe$_2$As$_2$, in the $k_z = 0$ plane, shown in the unfolded Brillouin zone (with one Fe per unit cell)[3]. It consists of two zone-centered hole pockets ($h_1$ and $h_2$) and one hole pocket at $M$ ($h_3$). **b)** Sketch of the main Fermi surface sheets of K$_{1-x}$Ba$_x$Fe$_2$As$_2$, at $x = 0.4$, with an electron pocket ($e$) at $X$ (and no hole pocket at $M$)[3]. The standard $s_\pm$ pairing state (type I) involves full gaps on each pocket, with a sign change from + on the hole pockets ($h_1$, $h_2$) to − on the electron pocket ($e$). **c)** $d$-wave pairing state[19], where the gap changes sign as the azimuthal angle crosses the zone diagonals (dashed lines). This symmetry forces the gap to have nodes (zeros) on all Fermi surface sheets that cut those diagonals (small open circles). **d)** Illustration of an $s_\pm$ pairing state (type II) where the gap changes sign from + on the inner $\Gamma$-centered hole pocket ($h_1$) to − on the outer $\Gamma$-centered hole pocket ($h_2$) (see ref. 15).

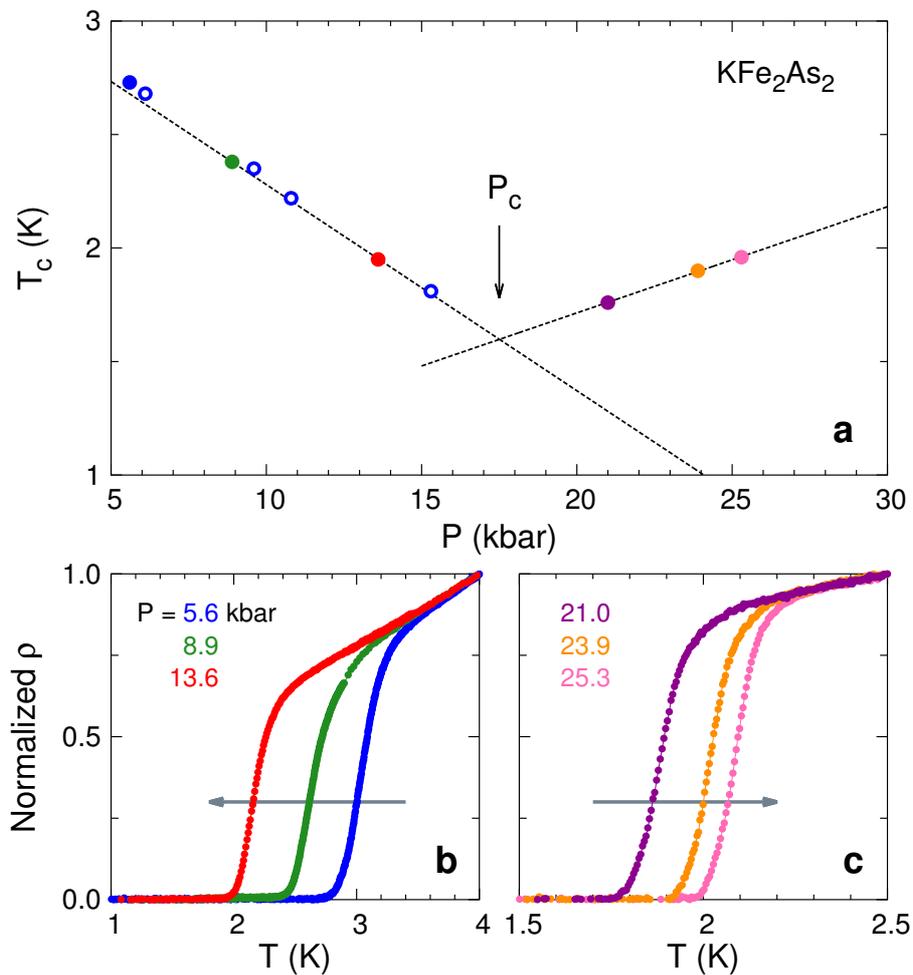

**Figure 2 | Pressure dependence of $T_c$ in KFe$_2$As$_2$.**

**a)** Pressure dependence of the superconducting transition temperature $T_c$ in KFe$_2$As$_2$. Full circles represent sample A (from color-coded resistivity data in panels b and c) and open circles sample B. The critical pressure $P_c$ = 17.5 kbar marks the transition from a decreasing to an increasing $T_c$. The dotted lines are linear fits to the data within 10 kbar on either side of $P_c$. **b)** Isobars of $\rho(T)$ in sample A, normalized to unity at $T$ = 4 K, for three pressures below $P_c$, as indicated. **c)** Same as in panel b, for pressures above $P_c$, with $\rho$ normalized at $T$ = 2.5 K. $T_c$ is the temperature below which $\rho(T)$ = 0. The accuracy on $T_c$ is better than ± 1% (Fig. S2). The grey arrows show how $T_c$ moves with pressure.

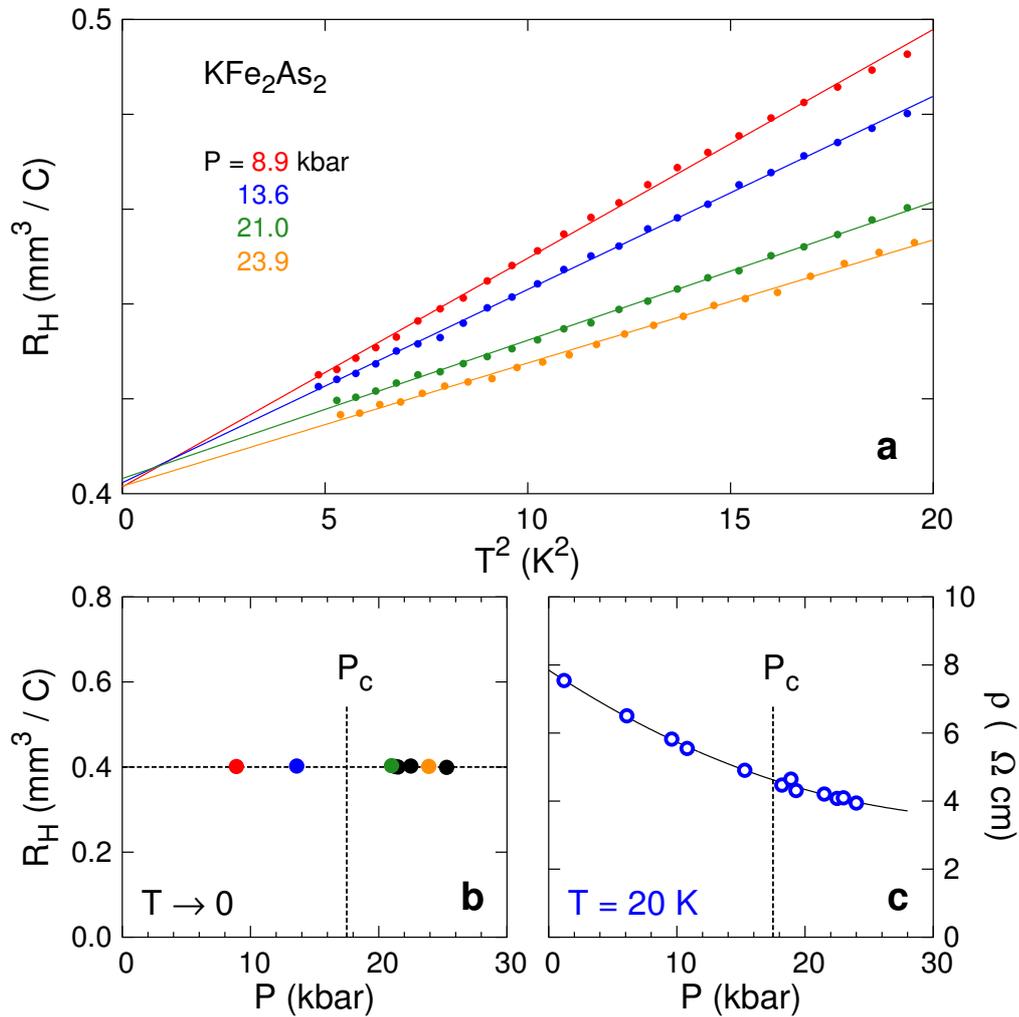

**Figure 3 | Hall coefficient and resistivity of KFe₂As₂ under pressure.**

**a)** Hall coefficient $R_H$ of KFe₂As₂ as a function of temperature, plotted as $R_H$ vs $T^2$, measured in sample A for a magnetic field of 13 T along the *c* axis of the tetragonal lattice, at different values of the applied pressure *P*, as indicated. The lines are linear fits to the data, extrapolated to *T* = 0 to yield the residual values, $R_H(0)$. **b)** Pressure dependence of $R_H(0)$, seen to remain completely unchanged throughout the range investigated. **c)** Pressure dependence of the resistivity $\rho$ measured in sample A at *T* = 20 K, seen to decrease monotonically through $P_c$. The solid line is a smooth fit through all data points.

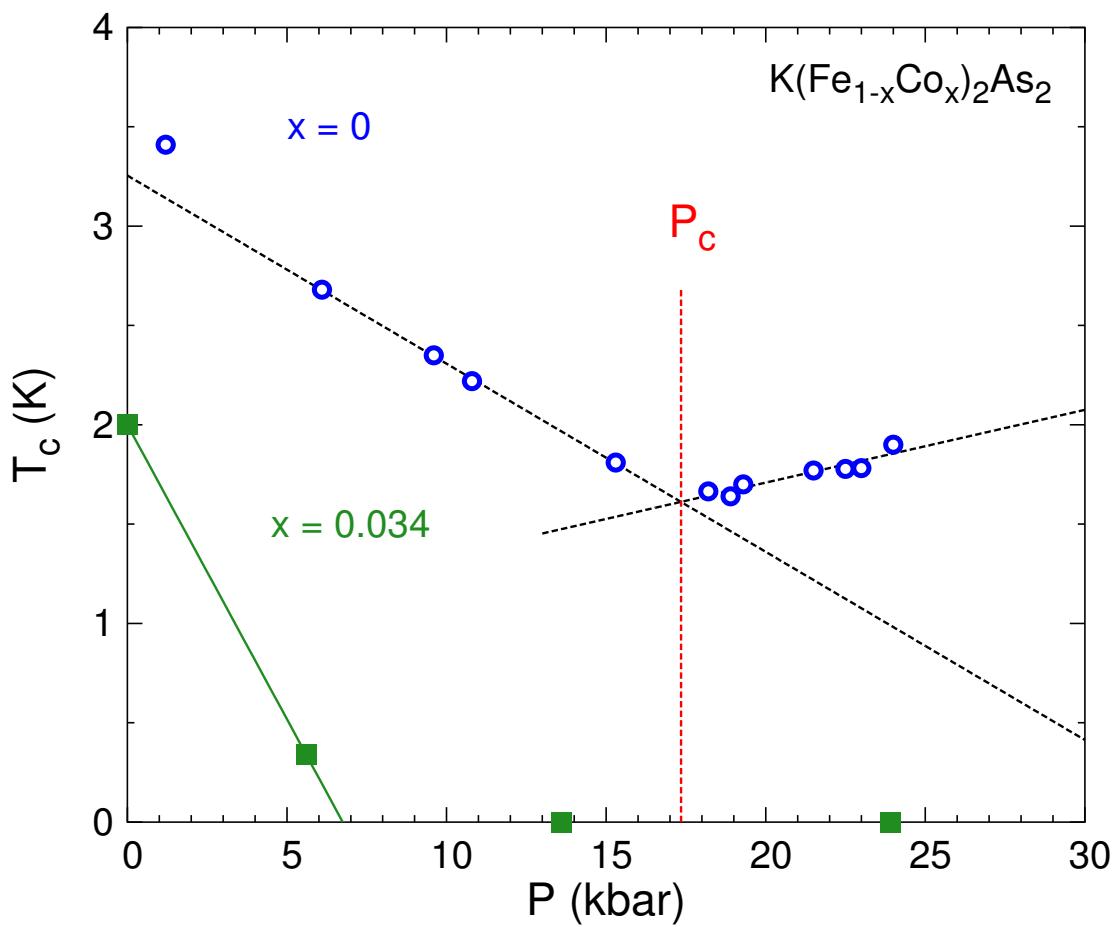

**Figure 4 | Effect of impurity scattering on $T_c$ in KFe$_2$As$_2$ .**

Pressure dependence of $T_c$ in KFe$_2$As$_2$ for a nominally pure sample (open circles, sample B) and a sample with 3.4% Co impurities (full squares, K(Fe$_{1-x}$Co$_x$)$_2$As$_2$ with $x$ = 0.034, from resistivity data in Fig. S3). The dotted lines are linear fits to the pure data within 10 kbar on either side of $P_c$. The solid green line is a linear fit to the first two data points of the impure sample. Note how $T_c$ is suppressed by the addition of impurities both below and above $P_c$, showing that the pairing states on both sides of the transition are highly sensitive to impurity scattering.

# SUPPLEMENTARY INFORMATION

## Change of pairing symmetry in the iron-based superconductor KFe$_2$As$_2$

### SAMPLES

Single crystals of KFe$_2$As$_2$ were obtained by KAs flux growth at USTC [13]. Three crystals of KFe$_2$As$_2$ were measured: two pure and one doped with 3.4% of Co, labeled A, B, and C, respectively. Table S1 lists the value of $T_c$, the residual resistivity $\rho_0$, and residual resistivity ratio RRR (= $\rho$(300 K) / $\rho_0$) for each sample, at ambient pressure. The superconducting transition temperature $T_c$ is defined as the temperature below which the resistance is zero.

| Sample | | $T_c$ (K) | $\rho_0$ (μΩ cm) | RRR |
|---|---|---|---|---|
| KFe$_2$As$_2$ | (A) | 3.54 | 0.25 | 1296 |
| KFe$_2$As$_2$ | (B) | 3.52 | 0.26 | 1246 |
| K(Fe$_{1-x}$Co$_x$)$_2$As$_2$ | (C) | 1.7 | 3.8 | 66 |

**Table S1 | Sample characteristics.**

Superconducting transition temperature ($T_c$), residual resistivity ($\rho_0$), and residual resistivity ratio RRR for the three samples studied in this work, all measured at ambient pressure. (Sample C: $x$ = 0.034.)

## METHODS

Resistivity ($\rho_{xx}$) and Hall resistance ($\rho_{xy}$) were measured under pressure in a six-probe AC configuration. Six contacts were soldered on each sample; one longitudinal pair to pass the electric current, another longitudinal pair to measure $\rho_{xx}$, and one transverse pair to measure $\rho_{xy}$. The samples were mounted so that the magnetic field $H$ was parallel to the tetragonal $c$ axis. Hall measurements were done by reversing the field and anti-symmetrizing the $\rho_{xy}$ data. At each pressure, the Hall signal was measured by sweeping temperature from 20 K to 2 K at $H$ = + 13 T and $H$ = - 13 T.

Samples were pressurized in a piston-cylinder clamp cell made of Be-Cu alloy 25 with the inner jacket made of alloy MP35N. The pressure inside the cell was determined via the $T_c$ of a lead wire. Measurements on the two pure samples (A and B) were repeated in two different pressure cells, with 4-mm and 6-mm inner jacket diameters, respectively. They were also performed in two different pressure media: Daphne oil 7373 and a 1/1 mixture of pentane and 3-methyl-1-butanol. As shown in Fig. S1, the values of $T_c$ vs $P$ obtained are all in excellent agreement, demonstrating that our results are highly reproducible and independent of the choice of sample, pressure medium or pressure cell. (They are also independent of the choice of $T_c$ criterion – see below.)

The resistivity under pressure was measured in the temperature range from 300 K to 0.3 K using a Cambridge Magnetic Refrigerator. A calibrated Cernox attached directly on the cell was used for thermometry. For measurements below 2 K, a calibrated ruthenium oxide thermometer mounted on the low-temperature stage was used. Slow cooling rates of 100 mK / min (between 300 K and 2 K) and 20 mK / min (between 2 K and 0.3 K) were used to maintain a good thermal equilibrium between the cell and the stage. Resistivity curves below 2 K were obtained in both cooling and heating cycles with no detectable thermal hysteresis.

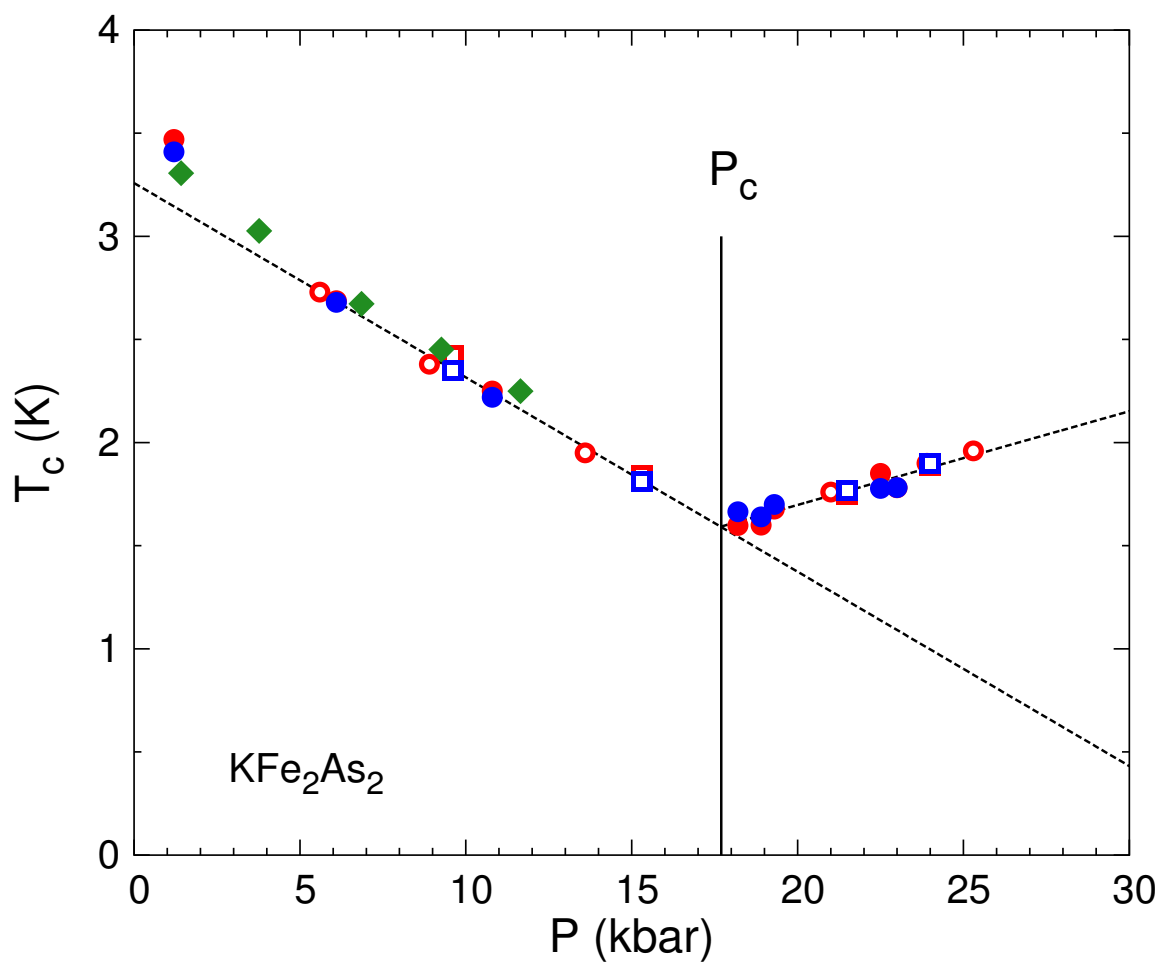

**Figure S1 | Comparing different measurements of $T_c$ vs pressure in KFe$_2$As$_2$.**

Superconducting critical temperature $T_c$ of KFe$_2$As$_2$, defined as the point of zero resistance, measured on our two pure samples (A, blue symbols; B, red symbols). Data obtained in two pressure cells (6-mm cell, circles; 4-mm cell, squares) and in two pressure media (Daphne oil, full symbols; pentane mixture, open symbols) are compared. Excellent reproducibility is observed, across different samples, cells and media. Our data are also in good agreement with those of the only other study of KFe$_2$As$_2$ under pressure (green diamonds, from ref. 29).

# DETERMINATION OF $T_c$

In this paper, the superconducting transition temperature $T_c$ is defined as the temperature below which the resistance is zero. We emphasize that in both samples A and B the width of the superconducting transition was narrow and, importantly, it remained constant as a function of pressure, as can be seen in Figs. 2b and 2c. This implies that the dependence of $T_c$ on pressure reported in Fig. 2a (from sample A) and Fig. 4 (from sample B) does not depend on our choice of criterion (*e.g.* whether it is $R = 0$ or 10%, 50%, 90% of the resistive drop), modulo a rigid offset. The best way to demonstrate this is to plot (normalized) $\rho$ vs $T / T_c$ for several isobars, as done in Fig. S2. This scaling analysis avoids having to choose a particular criterion for defining $T_c$ and affords greater accuracy, with an uncertainty of less than $\pm$ 1% on the relative value of $T_c$ (as it changes with pressure). In Fig. S2a, we see that $T_c$ varies strictly linearly with $P$ on both sides of $P_c$, within this very small error bar. This confirms that $T_c$ has a kink at $P_c$, the signature of a transition.

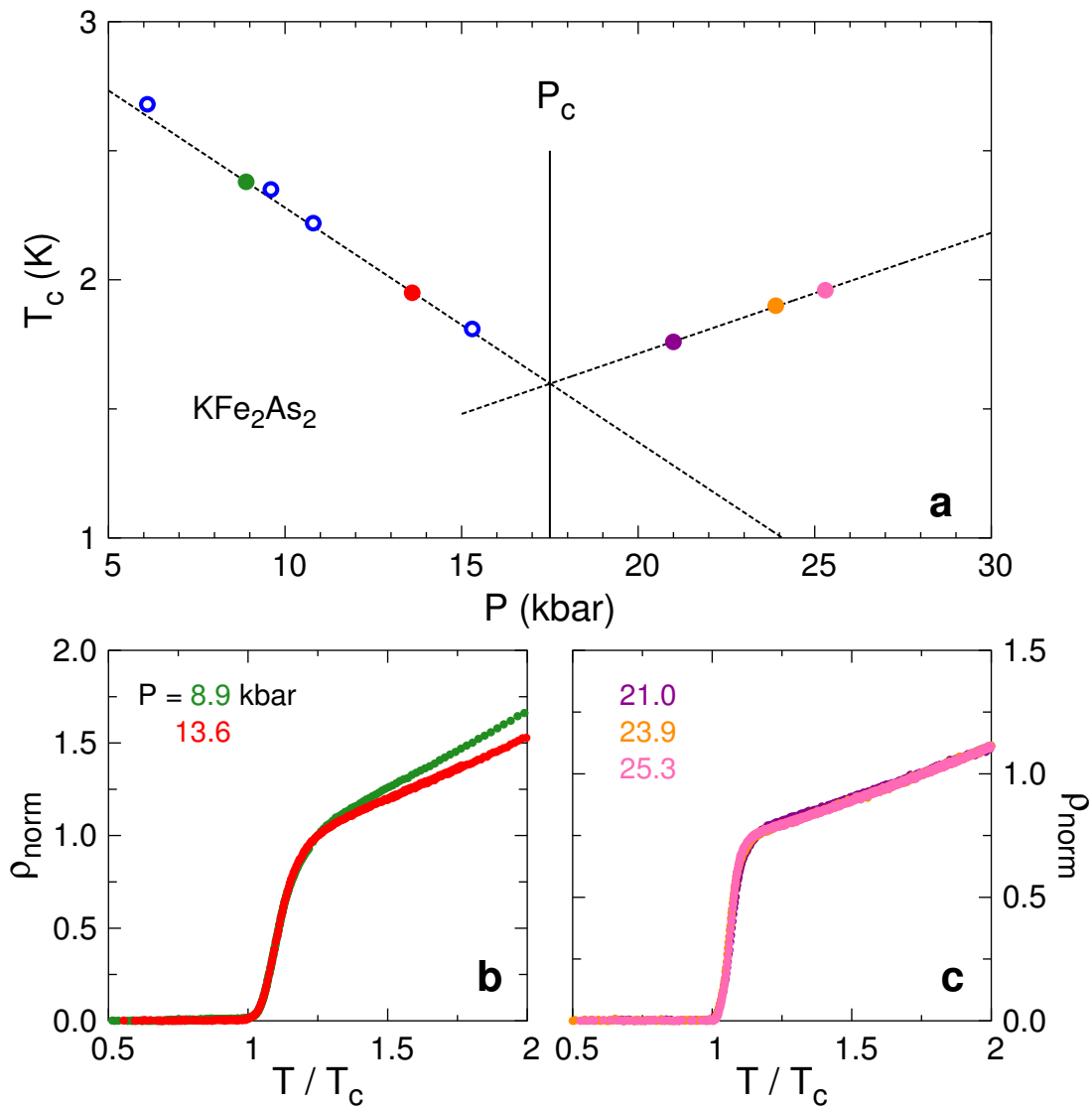

**Figure S2 | Determination of $T_c$ in KFe$_2$As$_2$.**

**a)** Pressure dependence of $T_c$ in sample A (full circles; from data in Fig. 2) and sample B (open circles; Fig. 4) obtained from isobars as in panel b (for $P < P_c$) and panel c (for $P > P_c$). The colour of full circles matches that of the corresponding isobars in panels b and c. The two dashed lines are a linear fit through the full circles on either side of $P_c$. **b)**, **c)** Isobars of $\rho(T)$ in sample A at different pressures as indicated, plotted as $\rho(T)$ vs $T / T_c$, where $\rho$ is normalized and $T_c$ is adjusted so that all curves collapse onto a single curve. The resulting $T_c$ values, accurate to better than ± 1%, are plotted in panel a and in Fig. 2a (as colour-coded full circles).

**IMPURITY SCATTERING IN KFe$_2$As$_2$**

One way to differentiate a *d*-wave superconductor from a standard *s*-wave superconductor (*s*$_{++}$) is to study the effect of non-magnetic impurities on $T_c$. In a *d*-wave superconductor, where the sign change in the order parameter around the Fermi surface is imposed by symmetry, impurity scattering rapidly suppresses $T_c$ because it mixes the phase of initial and final states. For a single-band system, the critical scattering rate $\Gamma_c$ for suppressing $T_c$ to zero is on the order of the clean limit $T_{c0}$ [30] :

$$\hbar\, \Gamma_c = 0.88\, k_B\, T_{c0}$$

This is indeed what is found in KFe$_2$As$_2$ [11, 12, 13], and this is one of several arguments in favour of a *d*-wave state in this iron arsenide at ambient pressure [11, 12]. The situation is very different in the iron-based superconductors CaFe$_2$As$_2$, BaFe$_2$As$_2$, and SrFe$_2$As$_2$ at optimal doping, where $\hbar\, \Gamma_c \sim 45\, k_B\, T_{c0}$ [27]. This robustness is consistent with an *s*$_{++}$ state. It can also be consistent with a type-I *s*$_{\pm}$ state (with sign change between hole and electron pockets; Fig. 1b) if inter-band impurity scattering is weak [3]).

The type-II *s*$_{\pm}$ state proposed by Maiti *et al.* [15] (with sign change between $h_1$ and $h_2$ hole pockets; Fig. 1d) is likely to be more sensitive to disorder than the type-I *s*$_{\pm}$ state. This is because for those two rather similar hole pockets, inter-band and intra-band impurity scattering are very likely to be comparable.

To determine the sensitivity of $T_c$ to impurity scattering in KFe$_2$As$_2$ as a function of pressure, the resistivity and Hall effect were measured in a sample of KFe$_2$As$_2$ with 3.4% Co impurities (sample C, Table S1). At ambient pressure, $T_c$ = 1.7 K [13]. As seen in Fig. S3, pressure rapidly reduces $T_c$ to values below 0.3 K, and superconductivity does not reappear at high pressure. This therefore rules out an *s*$_{++}$ state at $P > P_c$ (and a type-I *s*$_{\pm}$ state). As in the pure sample, the Hall coefficient $R_H(0)$ of the Co-doped sample does not show any change across $P_c$, confirming here as well that there is no Lifshitz transition.

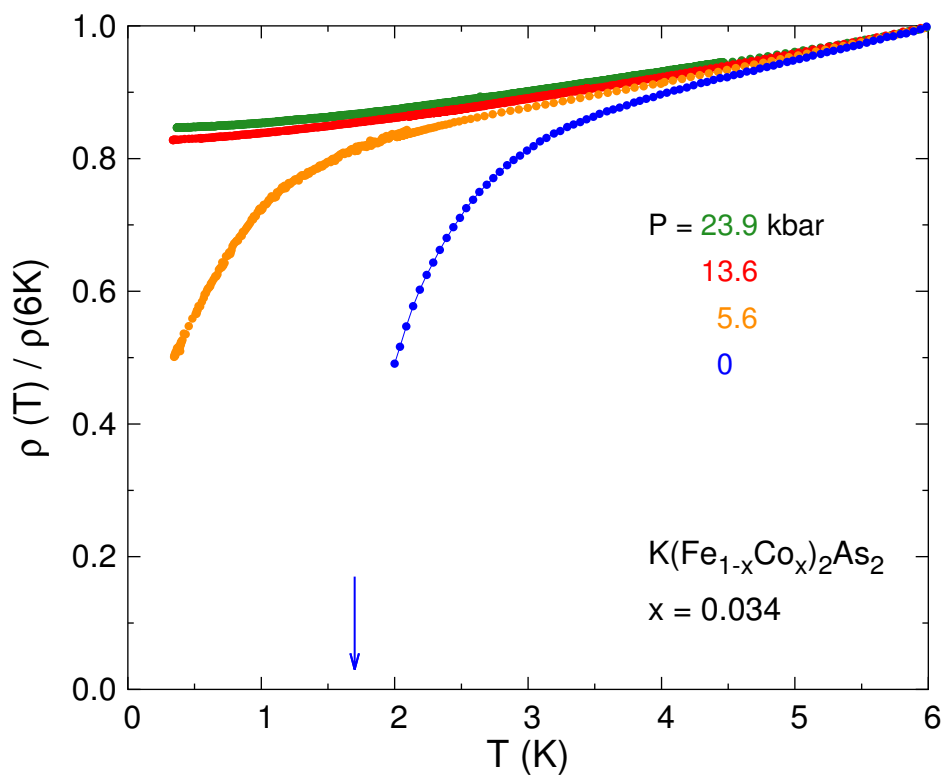

**Figure S3 | Resistivity of a KFe$_2$As$_2$ sample with Co impurities.**

Electrical resistivity of a sample of KFe$_2$As$_2$ with 3.4% Co impurities, normalized to its value at $T$ = 6 K, for different pressures as indicated. At ambient temperature, $T_c$ = 1.7 K [13] (arrow). At higher pressure, $T_c$ drops below 0.3 K and remains undetected up to the highest pressures. This shows that the superconducting states below and above $T_c$ are both easily destroyed by a low level of impurity scattering. The $T_c$ values displayed in Fig. 4 (as full green squares) are obtained using the criterion that $\rho$ has fallen to 50% of its value at $T$ = 6 K.

**PRIOR WORK**

We are aware of only one previous study of $KFe_2As_2$ under pressure, reported in ref. 29. Our $T_c$ values obtained from resistivity at different pressures are in good agreement with those of ref. 29 (Fig. S1), where $T_c$ is defined as the onset of the magnetization drop. Note that because the maximal pressure in that study was 12 kbar, the transition at $P_c$ was not observed.